\documentclass[aps,showpacs,tightenlines,nobalancelastpage,
floatfix,onecolumn]{revtex4}

\usepackage{amsmath}
\usepackage{amssymb}
\usepackage{graphicx}
\usepackage{dcolumn}
\usepackage{bm}
\usepackage{epic}

\begin{document}

\title{Optical pumping via incoherent Raman transitions}

\date{\today }

\author{A.~D. Boozer, R. Miller, T. E. Northup, A. Boca, and H.~J. Kimble}

\affiliation{
  Norman Bridge Laboratory of Physics 12-33,
  California Institute of Technology,
  Pasadena, CA 91125
}

\begin{abstract}
  A new optical pumping scheme is presented that uses incoherent Raman
  transitions to prepare a trapped Cesium atom in a specific Zeeman
  state within the $6S_{1/2}, F=3$ hyperfine manifold.
  An important advantage of this scheme over existing optical pumping
  schemes is that the atom can be prepared in any of the $F=3$ Zeeman
  states.
  We demonstrate the scheme in the context of cavity quantum
  electrodynamics, but the technique is equally applicable to a wide
  variety of atomic systems with hyperfine ground-state structure.
\end{abstract}


\maketitle

\section{Introduction}

Many experiments in atomic physics rely on the ability to prepare
atoms in specific internal states.
For example, spin-polarized alkali atoms can be used to polarize the
nuclei of noble gases \cite{Walker97}, to act as sensitive
magnetometers \cite{Budker02}, and to provide frequency standards
that exploit magnetic-field-insensitive clock transitions
\cite{Audoin92}.
In the field of quantum information science, internal atomic states
can be used to store and process quantum bits
\cite{Cirac97,Zoller05,Laurant,Chou07,Leibfriend03} with extended
coherence times.

A standard method for preparing an atom in a specific internal state
is optical pumping \cite{Kastler66,Demtroder82,Happer72}, which
involves driving the atom with light fields that couple to all but one
of its internal states; these light fields randomly scatter the atom
from one internal state to another until it falls into the uncoupled
``dark'' state.
Various optical pumping schemes have been analyzed and demonstrated
for alkali atoms \cite{Cutler80, Audoin92, Wang07} and today are
well-established techniques.
These schemes rely on dark states that are set by the polarization of
the driving field, and this imposes restrictions on the possible
Zeeman states in which the atom can be prepared.
Specifically, one can prepare the atom in the $m_F = 0$ state by using
light that is linearly polarized along the
quantization axis, or in one of the edge states ($m_F=\pm F$) by using
light that is circularly $\sigma_\pm$-polarized
along the quantization axis.

In contrast, the scheme presented here allows the atom to be prepared
in any of the Zeeman states within the lowest ground state hyperfine
manifold of an alkali atom, which in our case is the $6S_{1/2}, F=3$
manifold of Cesium.
The key component of the scheme is a pair of optical fields that drive
Raman transitions between pairs of Zeeman states
$|3,m\rangle \leftrightarrow |4,m\rangle$.
We apply a magnetic bias field to split out the individual Zeeman
transitions, and add broadband noise to one of the optical fields,
where the spectrum of the noise is tailored such that all but one of
the transitions are driven.
The two Zeeman states corresponding to the undriven transition
are the dark states of the system, and we exploit these dark states
to perform optical pumping.
We verify the optical pumping by using coherent Raman transitions to
map out a Raman spectrum, which allows us to determine how the atomic
population is distributed among the different Zeeman states.
The capability of driving Raman transitions between hyperfine ground
states has many additional applications, such as state manipulation
\cite{Wineland03},
ground state cooling
\cite{Monroe95,Hamann98,Vuletic98,Boozer06},
precision measurements
\cite{Clade06,Gustavson97},
and Raman spectroscopy
\cite{Dotesenko04}.
The scheme described here shows that this versatile tool can also be
used for atomic state preparation.

We have demonstrated this scheme in the context of cavity quantum
electrodynamics (QED), specifically in a system in which a single
atom is strongly coupled to a high-finesse optical cavity.
Cavity QED offers a powerful resource for quantum information science,
and the ability to prepare the atom in a well-defined initial state
is a key requirement for many of the protocols that have been proposed
for this system, such as the generation of polarized single photons
\cite{Birnbaum-thesis, Wilk07a} and the
transfer of Zeeman coherence to photons within the cavity mode
\cite{Parkins93}.
Conventional optical pumping to a single Zeeman sublevel has been
previously demonstrated within a cavity \cite{Wilk07b}, but we find
our new method to be particularly effective given the constraints of
our system, in which optical access to the atom is limited and we must
address the multiplicity of Cesium sublevels.
However, optical pumping via incoherent Raman
transitions has much broader applications beyond the cavity QED
setting, and can be used in a wide variety of atomic systems with
hyperfine ground-state structure.

\section{Experimental apparatus}
\label{experimental-apparatus}

\begin{figure}
  \centering
  \includegraphics[width=12cm]{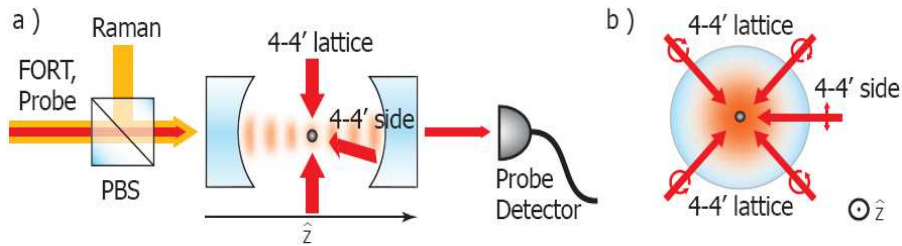}
  \caption{
    Schematic of experiment.
    (a) View from the side of the cavity.
    Shown are the linearly polarized FORT, Raman, and probe beams that
    drive the cavity, and the circularly polarized $4-4'$ lattice
    beams and linearly polarized $4-4'$ side beam that drive the atom.
    (b) View along the cavity axis.
    Shown are the $4-4'$ lattice beams and the $4-4'$ side beam.
  }
  \label{fig:experiment-schematic}
\end{figure}

Our system consists of a single Cesium atom that is strongly coupled
to a high-finesse optical cavity, as shown in Figure
\ref{fig:experiment-schematic}.
The cavity supports a set of discrete modes, and its length is tuned
so that one pair of modes \footnote{
  Since there are two polarization degrees of freedom, the cavity
  modes occur in nearly-degenerate pairs.
} is nearly resonant with the atomic transition
$6S_{1/2}, F=4 \rightarrow 6P_{3/2}, F=5'$ at $\lambda_{D2} =
852\,\mathrm{nm}$.
The atomic dipole associated with this transition
couples to the electric field of the resonant mode, allowing the
atom and cavity to exchange excitation at a characteristic rate
$g = (2\pi)(34\,\mathrm{MHz})$ for the
$6S_{1/2}, F=4, m_F=4 \rightarrow 6P_{3/2}, F=5',m_{F'}=5$ transition,
a rate that is much larger than either the
cavity decay rate $\kappa = (2\pi)(3.8\,\mathrm{MHz})$ or the atomic
decay rate $\gamma = (2\pi)(2.6\,\mathrm{MHz})$; thus, the system is
in the strong-coupling regime \cite{Miller05}.

We hold the atom inside the cavity via a state-insensitive
far off-resonance trap (FORT) \cite{McKeever03}.
The FORT is produced by resonantly driving a cavity mode at
$\lambda_F = 936\,\mathrm{nm}$ with a linearly polarized beam,
which creates a red-detuned standing wave inside the cavity.
Each antinode of this standing wave forms a potential well in which an
atom can be trapped; for the experiments described here, the optical
power of the FORT beam is chosen such that the depth of these wells is
$U_F = (2\pi)(45\,\mathrm{MHz})$.

We drive Raman transitions between the $F=3$ and $F=4$ hyperfine
ground-state manifolds of the atom by adding a second beam, referred
to here as the Raman beam, which drives the same cavity mode as the
FORT beam but is detuned from the FORT  by the atomic hyperfine splitting
$\Delta_{HF} = (2\pi)(9.2\,\mathrm{GHz})$
(this scheme was first proposed in \cite{Boozer-thesis}, and was used
to perform Raman sideband cooling in \cite{Boca}).
The FORT and Raman beams are combined on a polarizing beam splitter
(PBS) before entering the cavity, so the Raman beam is linearly
polarized in a direction orthogonal to the polarization of the FORT
beam.
To stabilize the frequency difference between the FORT and Raman
beams, the external-cavity diode laser that generates the Raman beam
is injection-locked to the red sideband of light that has
been picked off from the FORT beam and passed through an
electro-optical modulator (EOM), which is driven at $\Delta_{HF}$.
The FORT and Raman beams form the two legs of a Raman pair and drive
Raman transitions between pairs of Zeeman states
$|3,m\rangle \leftrightarrow |4,m\rangle$, where the quantization axis
$\hat{z}$ is chosen to lie along the cavity axis \footnote{
  The FORT-Raman pair generates a Raman coupling between the hyperfine
  ground states that is proportional to
  $\vec{J}\cdot(\hat{\epsilon}_F \times \hat{\epsilon}_R)$, where
  $\vec{J}$ is the electron angular momentum operator and
  $\hat{\epsilon}_F$,$\hat{\epsilon}_R$ are the polarization vectors
  for the FORT and Raman beams, so in general $\Delta m = \pm 1, 0$
  transitions are possible \cite{Boozer-thesis}.
  For our system $\hat{\epsilon}_F \times \hat{\epsilon}_R = \hat{z}$,
  so only the $\Delta m = 0$ transitions are driven.
}.
Typically we use a strong FORT beam and a weak Raman beam, so the
Raman beam does not significantly alter the FORT trapping potential
\footnote{
  The FORT and Raman beams give level shifts
  $U_F \sim \Omega_F^2/\Delta$ and $U_R \sim \Omega_R^2/\Delta$,
  and the effective Rabi frequency for the Raman transitions driven by
  the FORT-Raman pair is $\Omega_E \sim \Omega_F \Omega_R/\Delta$, where
  $\Omega_{F,R}$ are the Rabi frequencies of the FORT and Raman beams
  and $\Delta$ is the detuning from atomic resonance.
  Thus, the ratio of the level shifts is
  $U_R/U_F \sim (\Omega_E/U_F)^2 \sim 10^{-5}$ for the typical values
  $U_F = (2\pi)(45\,\mathrm{MHz})$,
  $\Omega_E = (2\pi)(120\,\mathrm{kHz})$.
}.

In order to address individual Zeeman transitions, we apply a
magnetic bias field $B_a$ along the cavity axis.
This axial field shifts the
$|3,m\rangle \leftrightarrow |4,m\rangle$ transition by
\begin{eqnarray}
  \label{eqn:zeeman-shifts}
  \delta(|3,m\rangle \leftrightarrow |4,m\rangle) =
  \omega_B\,m,
\end{eqnarray}
where
\begin{eqnarray}
  \omega_B \equiv (g_4 - g_3) \mu_B B_a =
  (2\pi)(700\,\mathrm{kHz/G})\,B_a,
\end{eqnarray}
and $g_4 = 1/4$, $g_3 = -1/4$ are the Lande $g$-factors for the $F=4$
and $F=3$ ground-state hyperfine manifolds. For the experiments
described here, we typically set the axial bias field such that
$\omega_B \simeq (2\pi)(910\,\mathrm{kHz})$.

The strong atom-cavity coupling allows us to to determine whether the
atom is in the $F=3$ or $F=4$ hyperfine manifold by driving the cavity
with a $100\,\mathrm{\mu s}$ pulse of resonant $4-5'$
probe light, as described in \cite{Boozer06}.
If the atom is in $F=4$, it couples to the cavity and blocks the
transmission of the probe beam, while if the atom is in $F=3$, it
decouples from the cavity, and the probe beam is transmitted.
Using this technique, we can determine the hyperfine ground state
of the atom with an accuracy of $\sim 98\%$ for a single
$100\,\mathrm{\mu s}$ measurement interval.

Atoms are delivered to the cavity by releasing a
magneto-optical trap located a few millimeters above the cavity, and
the falling atoms are loaded into the FORT by cooling them with $4-4'$
lattice light.
This lattice light consists of two pairs of counter-propagating beams
in the $\sigma_+-\sigma_-$ configuration, which are applied from the
sides of the cavity.
We ensure that only one atom is trapped in the FORT by applying the
Raman beam and driving the cavity with a resonant $4-5'$ probe; this
combination gives an effect analogous to that in \cite{McKeever04},
which allows us to determine the number of atoms in the cavity based on
the amount of $4-5'$ light that is transmitted.

\section{Coherent and Incoherent Raman transitions}
\label{sec:coherent-and-incoherent}

If the FORT and Raman beams are both monochromatic, then they drive
coherent Raman transitions between pairs of Zeeman states
$|3,m\rangle \leftrightarrow |4,m\rangle$, and the atomic
populations oscillate between the two states in each pair.
The effective Rabi frequency for the
$|3,m\rangle \leftrightarrow |4,m\rangle$ transition is
\begin{eqnarray}
  \Omega_E(|3,m\rangle \leftrightarrow |4,m\rangle) =
  \Omega_0\,(1 - m^2/16)^{1/2},
\end{eqnarray}
where $\Omega_0$ is set by the power in the FORT and Raman beams
\cite{Boozer-thesis}.
For the experiments described here, the powers in these beams are
chosen such that that $\Omega_0 \simeq (2\pi)(120\,\mathrm{kHz})$.
The Raman detuning for the FORT-Raman pair is given by
$\delta_R = \omega_F - \omega_R - \Delta_{HF}$,
where $\omega_F$ and $\omega_R$ are the optical frequencies of the
FORT and Raman beams, which means that the effective detuning for the
$|3,m\rangle \leftrightarrow |4,m\rangle$ transition is
\begin{eqnarray}
  \delta_E(|3,m\rangle \leftrightarrow |4,m\rangle) =
  \delta_R - \omega_B\,m.
\end{eqnarray}
We can also drive incoherent Raman transitions by using a
monochromatic FORT beam and a spectrally broad Raman beam, where the
spectral width is typically $\sim 10\,\mathrm{MHz}$.
In contrast to coherent Raman transitions, in which the atom undergoes
coherent Rabi oscillations, for incoherent Raman transitions the
atomic population decays at a constant rate from $|3,m\rangle
\rightarrow |4,m\rangle$ and from
$|4,m\rangle \rightarrow |3,m\rangle$.
In appendix \ref{sec:calculate-rate}, we show that these decay rates
are proportional to $S(\Delta_{HF} + \omega_B\,m)$, where $S(\omega)$
is the power spectrum of a beat note formed between the FORT and Raman
beams.

\section{Measuring the population distribution}
\label{measure-zeeman-populations}

Given an initial state of the atom in which the entire population lies
in the $F=3$ manifold, we can use coherent Raman transitions to
determine how the population is distributed among the various Zeeman
states.
To perform this measurement we prepare the atom in the desired initial
state, apply a coherent Raman pulse of fixed duration, Rabi frequency,
and Raman detuning, and then drive the cavity with a resonant
$F=4 \rightarrow F=5'$ probe beam to determine if the atom was
transfered to $F=4$.
By iterating this process we determine the probability $p_4$ for the
atom to be transfered by the Raman pulse, and by repeating the
probability measurement for different Raman detunings $\delta_R$ we
can map out a Raman spectrum $p_4(\delta_R)$.
For the Raman spectra presented here, the Raman pulses have Rabi
frequency $\Omega_0 = (2\pi)(120\,\mathrm{kHz})$ and duration
$25\,\mathrm{\mu s}$.
This is long enough that the Rabi oscillations decohere, and
the Raman spectrum just records the Lorentzian envelope for each Zeeman
transition.
Thus, when the $|3,m\rangle \leftrightarrow |4,m\rangle$ Zeeman
transition is resonantly driven by the Raman pulse, roughly half the
population that was initially in $|3,m\rangle$ is transfered to
$|4,m\rangle$.

\begin{figure}
  \centering
  \includegraphics[scale=0.5]{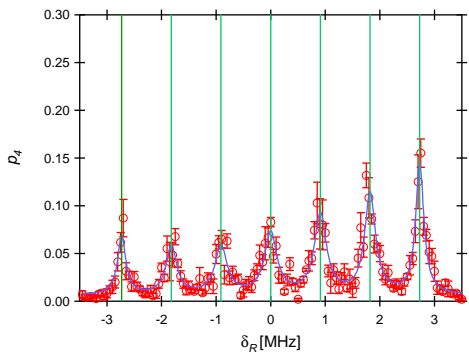}
  \caption{
    Raman spectrum for a random initial state.
    Shown is the transfer probability $p_4$ versus Raman detuning
    $\delta_R$:
    the points are the experimental data, the curve is a fit of
    $p_4(\delta_R)$, as given by equation (\ref{eqn:p4-deltaR}), and
    the vertical green lines indicate the predicted frequencies
    $\delta(|3,m\rangle \leftrightarrow |4,m\rangle)$ for individual
    Zeeman transitions.
  }
  \label{fig:scan-without-pumping}
\end{figure}

As a demonstration of this technique,
Figure \ref{fig:scan-without-pumping} shows a Raman
spectrum for an initial state with comparable populations in all
of the $F=3$ Zeeman states.
To prepare this state, we optically pump the atom to $F=3$ by
alternating $7$ pulses of resonant $F=4 \rightarrow F=4'$
lattice light with $7$ pulses of resonant $F=4 \rightarrow F=4'$
side light, where each pulse is $300\,\mathrm{ns}$ long.
The beams that deliver the lattice and side light are shown in Figure
\ref{fig:experiment-schematic}.

To determine the population $p_{3,m}$ in the Zeeman state
$|3,m\rangle$, we fit a sum of Lorentzians, one for each Zeeman
transition, to the experimental data:
\begin{eqnarray}
  \label{eqn:p4-deltaR}
  p_4(\delta_R) =
  p_b + (1/2)\sum_m
  (1 + (\delta_R - \omega_B\,m)^2/
  (1 - m^2/16)\,\Omega_0^2)^{-1}\,p_{3,m},
\end{eqnarray}
where $p_b$ is a constant background.
We fit the Zeeman state populations, the Rabi frequency $\Omega_0$,
and the frequency $\omega_B$ that characterizes the strength of the
axial bias field, and perform an independent measurement to determine
the background probability $p_b = 0.006$.
The fitted value of $\Omega_0$ agrees to within $14\,\%$ with the
value we would expect based on the measured optical powers in the FORT
and Raman beams, and the fitted value of $\omega_B$ agrees to within
$5\,\%$ with the value we would expect based on the known axial coil
current and geometry.
As a consistency check we sum the fitted populations and obtain
the result $1.10 \pm 0.03$, in reasonable agreement with the expected
value of $1$.

\section{Optical pumping scheme}
\label{optpumpscheme}

\begin{figure}
  \centering
  \includegraphics[scale=0.5]{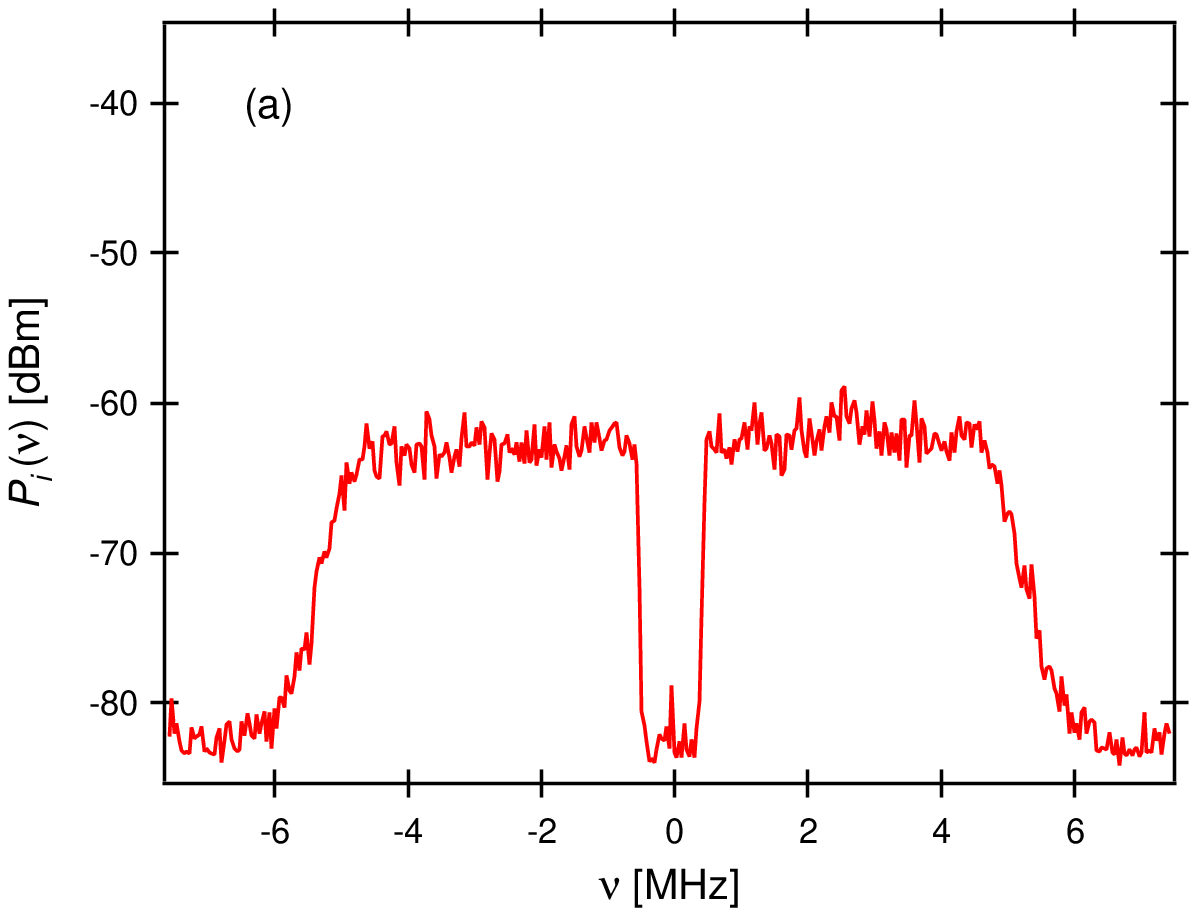}
  \includegraphics[scale=0.5]{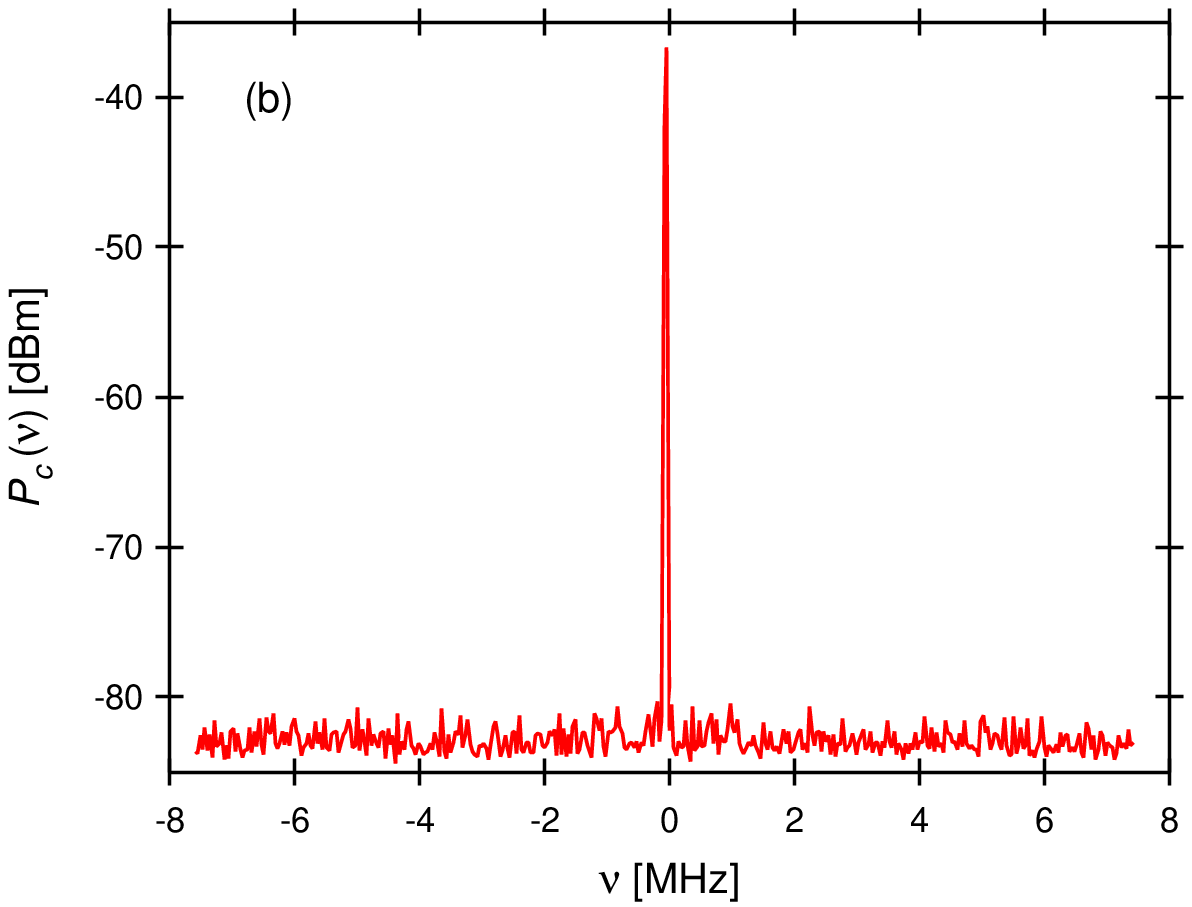}
  \caption{
    (a) Power spectrum of noise used for pumping into $|3,0\rangle$.
    (b) Power spectrum of coherent signal used for driving coherent
    Raman transitions with $\Omega_0 = (2\pi)(120\,\mathrm{kHz})$.
    Both curves are obtained by combining the FORT and Raman beams on
    a photodetector and measuring the spectrum of the photocurrent;
    shown is the RF power in a $3\,\mathrm{kHz}$
    bandwidth versus detuning from $\Delta_{HF}$.
  }
  \label{fig:noise-spectrum}
\end{figure}

We can prepare the atom in a specific Zeeman state by using a Raman
beam whose spectrum is tailored to incoherently drive all but one of
the Zeeman transitions.
As an example, Figure \ref{fig:noise-spectrum}a shows the power
spectrum of the noise used for pumping into $|3,0\rangle$.
This graph was obtained by measuring the power spectrum of a beat note
formed between the FORT and Raman beams by mixing them on a
photodetector with a non-polarizing beam splitter.
For comparison, Figure \ref{fig:noise-spectrum}b shows the power
spectrum for a monochromatic Raman beam tuned to Raman resonance, as
would be used for driving coherent Raman transitions.

Comparing the noise spectrum shown in Figure
\ref{fig:noise-spectrum}a to the Raman spectrum shown in Figure
\ref{fig:scan-without-pumping}, we see that the noise drives
incoherent Raman transitions from $|3,m\rangle \leftrightarrow
|4,m\rangle$ for $m \neq 0$, but because of the notch around zero
detuning, the $|3,0\rangle \leftrightarrow |4,0\rangle$ transition is
not driven.
We optically pump the atom into $|3,0\rangle$ by first driving
incoherent Raman transitions for $10\,\mathrm{\mu s}$, then pumping the
atom to $F=3$ using the procedure discussed in section
\ref{measure-zeeman-populations}, and iterating this sequence $40$
times.
It is straightforward to modify this procedure so as to pump into the
$|3,m\rangle$ Zeeman state for any $m$; we simply shift the notch in
the noise so that it overlaps with the $|3,m\rangle \leftrightarrow
|4,m\rangle$ transition.

\begin{figure}
  \centering
  \includegraphics[scale=0.5]{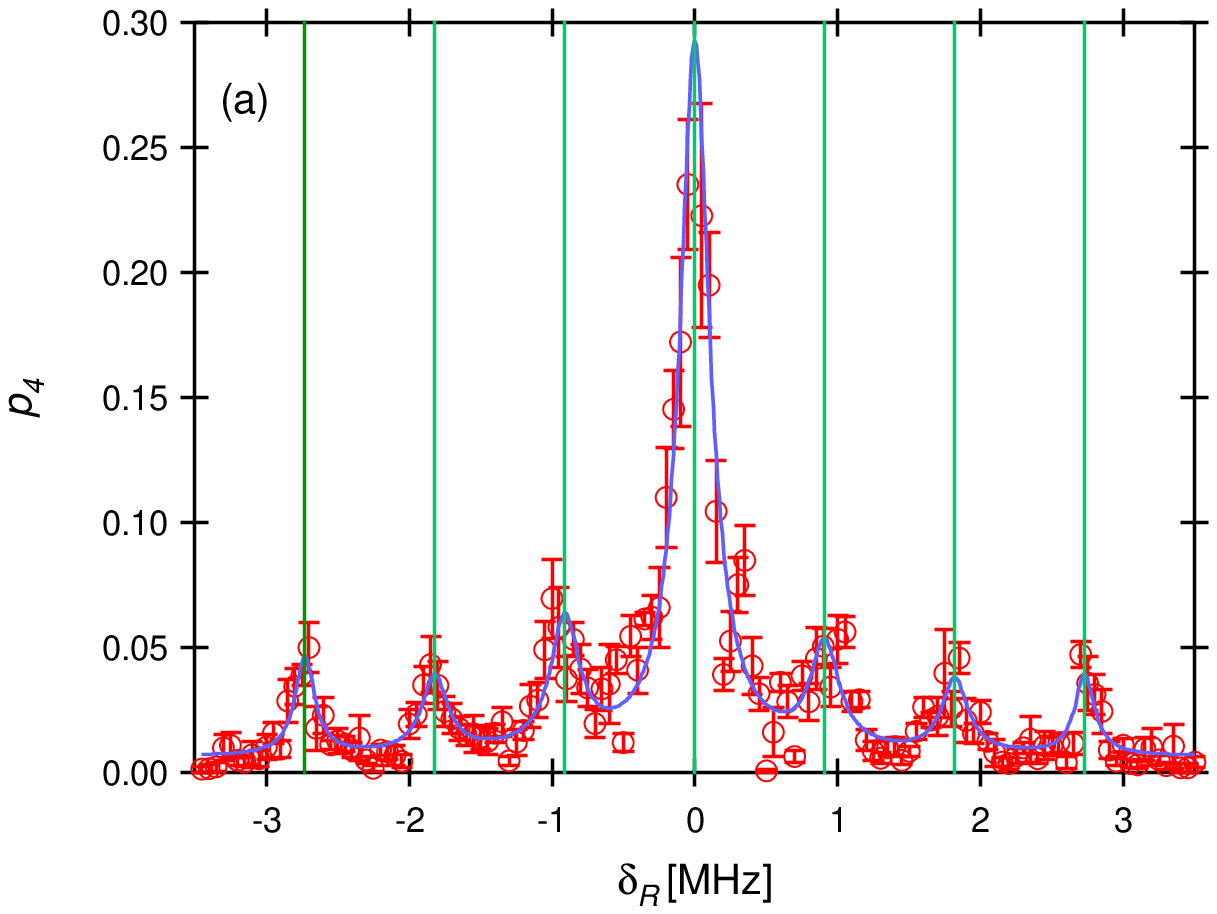}
  \includegraphics[scale=0.5]{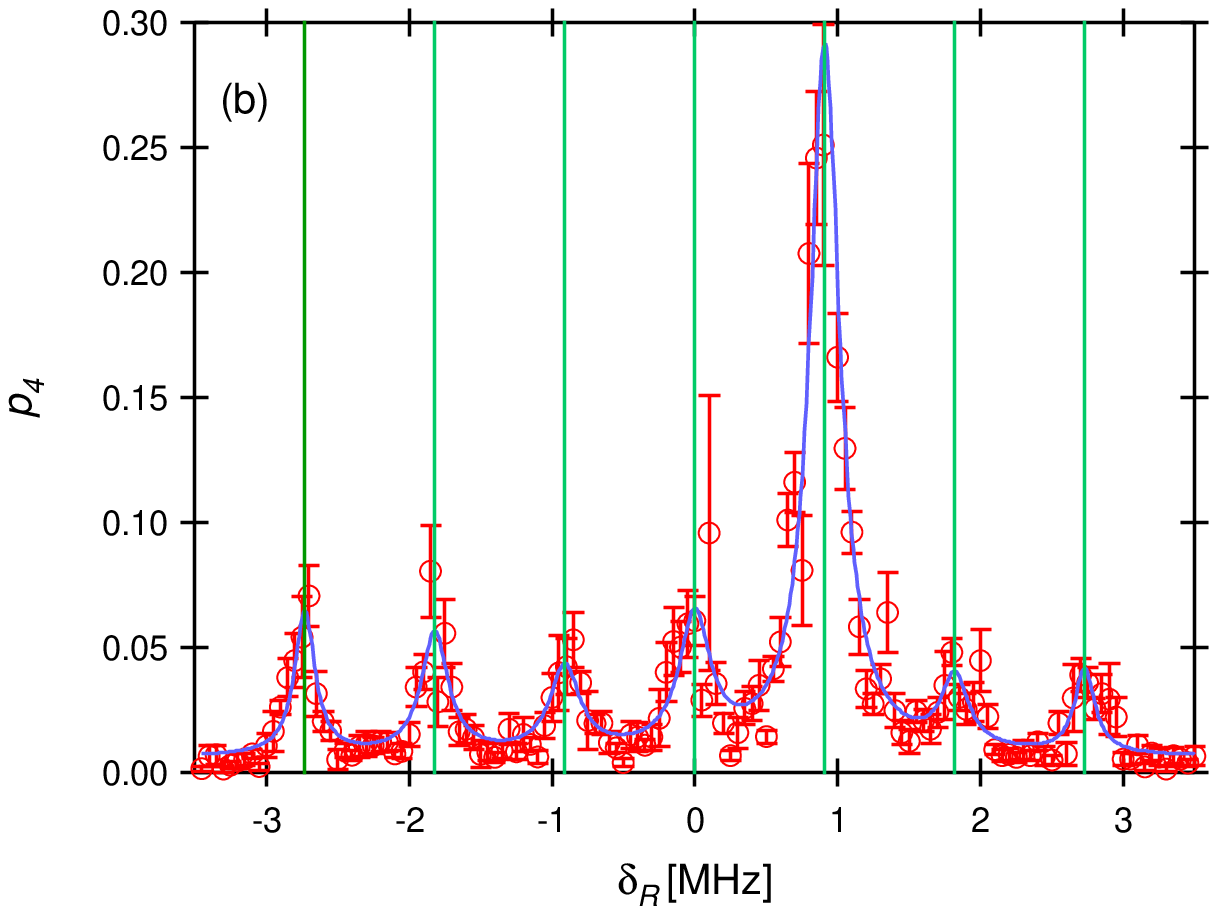}
  \caption{
    (a) Raman spectrum for optical pumping into $|3,0\rangle$.
    (b) Raman spectrum for optical pumping into $|3,1\rangle$.
    Raman spectrum for a random initial state.
    Shown is the transfer probability $p_4$ versus Raman detuning
    $\delta_R$:
    the points are the experimental data, the curve is a fit of
    $p_4(\delta_R)$, as given by equation (\ref{eqn:p4-deltaR}), and
    the vertical green lines indicate the predicted frequencies
    $\delta(|3,m\rangle \leftrightarrow |4,m\rangle)$ for individual
    Zeeman transitions.
  }
  \label{fig:incoherent-raman-pumping}
\end{figure}

To characterize the optical pumping, we first pump the atom into a
specific Zeeman state and then measure the Raman spectrum as described
in the preceding section.
Figure \ref{fig:incoherent-raman-pumping} shows Raman spectra
measured after pumping into (a) $|3,0\rangle$ and (b) $|3,1\rangle$.
We find that the fraction of the atomic population in the desired state
is $0.57 \pm 0.02$ for pumping into $|3,0\rangle$ and $0.57 \pm 0.02$
for pumping into $|3,1\rangle$, where the remaining population is
roughly equally distributed among the other Zeeman states
(these numbers are obtained using the by fitting equation
(\ref{eqn:p4-deltaR}) to the data, as described in section
\ref{measure-zeeman-populations}).
Summing the fitted populations in all the Zeeman states, we obtain the
value $1.02 \pm 0.04$ for (a) and $1.08 \pm 0.04$ for (b), in
reasonable agreement with the expected value of $1$.

To generate the Raman beam used in Figure \ref{fig:noise-spectrum}a,
we start with an RF noise source, which produces broadband noise that
is spectrally flat from DC to $\sim 10\,\mathrm{MHz}$.
The noise is passed through a high-pass filter at
$500\,\mathrm{kHz}$ and a low-pass filter at $5\,\mathrm{MHz}$,
where both filters roll off at $60\,\mathrm{dB}$ per octave.
The filtered noise is then mixed against an $85\,\mathrm{MHz}$
local oscillator, and the resulting RF signal is used to drive an
acousto-optical modulator (AOM) that modulates a coherent
beam from the injection-locked Raman laser.
The first order diffracted beam from the AOM forms a Raman beam with
the desired optical spectrum.
Note that previous work has demonstrated the use of both synthesized
incoherent laser fields \cite{Anderson90,Dinse88}, such as that used
here, as well as the noise intrinsic to free-running diode lasers
\cite{Lathi96,Yabuzaki91} to resonantly probe atomic spectra.

Although the scheme presented here relies on incoherent Raman
transitions, it is also possible to perform optical pumping with
coherent Raman transitions.
The basic principle is the same: we simultaneously drive all but one
of the Zeeman transitions, only instead of using a spectrally broad
Raman beam, we use six monochromatic Raman beams, where each beam is
tuned so as to resonantly drive a different transition.
We have implemented such a scheme, and found that it gives comparable
results to the incoherent scheme described above, but there are two
advantages to the incoherent scheme.
First, it is simpler to generate a Raman beam with the necessary
spectral properties for the incoherent scheme.
Second, when coherent Raman transitions are used, the six frequency
components for the Raman beam must be tuned to resonance with their
respective transitions, and hence are sensitive to the value of the
axial magnetic field.
When incoherent Raman transitions are used, however, the
same Raman beam can be used for a broad range of axial field values.

\section{Conclusion}

We have proposed a new scheme for optically pumping atoms into a
specific Zeeman state and have experimentally implemented the scheme
with Cesium atoms in a cavity QED setting.
An important advantage over existing schemes is that atoms can be
prepared in any of the Zeeman states in the lower hyperfine ground
state manifold.

We have measured the effectiveness of the optical pumping, and have
shown that a fraction $\sim 0.57$ of the atomic population can be
prepared in the desired Zeeman state.
Some possible factors that could be limiting the effectiveness of the
optical pumping include fluctuating magnetic fields transverse to the
cavity axis, misalignment of the cavity axis with the axial bias
field, and slow leaking out of the dark state due to scattering from
background light.
We are currently investigating these factors.

The scheme presented here operates on a fundamentally different
principle from existing optical pumping schemes, in that it relies on
incoherent Raman transitions to create an atomic dark state.
Raman transitions have many different applications in atomic physics,
so there are often independent reasons for incorporating a system for
driving Raman transitions into an atomic physics laboratory;
our scheme shows that such a system can also be applied to the problem
of atomic state preparation.
The scheme should serve as a useful tool for experiments in atomic
physics, both in a cavity QED setting and beyond.

This research is supported by the National Science Foundation,
the Army Research Office, and the Disruptive Technology Office of the
Department of National Intelligence.

\appendix

\section{Transition rate for incoherent Raman transitions}
\label{sec:calculate-rate}

As described in section \ref{sec:coherent-and-incoherent},
we drive incoherent Raman transitions between pairs of Zeeman
states $|3,m\rangle \leftrightarrow |4,m\rangle$ by using a
monochromatic FORT beam and a spectrally broad Raman beam.
For incoherent Raman transitions the atomic population decays at a
constant rate from $|3,m\rangle \rightarrow |4,m\rangle$ and from
$|4,m\rangle \rightarrow |3,m\rangle$, and in this appendix we
calculate these decay rates.

We will consider a single Zeeman transition
$|3,m\rangle \leftrightarrow |4,m\rangle$, so we can treat the system
as an effective two-level atom with ground state $g \equiv
|3,m\rangle$ and excited state $e \equiv |4,m\rangle$, where the
energy splitting between $g$ and $e$ is
$\omega_A \equiv \Delta_{HF} + \omega_B\,m$.
The FORT-Raman pair drives this effective two-level atom with broadband
noise, which we can approximate as a comb of classical fields with
optical frequencies $\omega_k$ and Rabi frequencies $\Omega_k$.
Let us assume that we start in the ground state $g$.
If we only consider the coupling of the atom to field $k$, then the
equation of motion for the excited state amplitude $c_e$ is
\begin{eqnarray}
  \label{eqn-of-motion}
  i\dot{c}_e = \frac{\Omega_k}{2}\,e^{-i\delta_k t}\,c_g,
\end{eqnarray}
where $\delta_k \equiv \omega_k - \omega_A$ is the detuning of the
field from the atom.
At small times the population is almost entirely in the ground state,
so we can make the approximation $c_g = 1$ and integrate equation
(\ref{eqn-of-motion}) to obtain
\begin{eqnarray}
  c_e(t) = \frac{\Omega_k}{2 \delta_k}(e^{-i\delta_k t} - 1).
\end{eqnarray}
Thus, the transition rate from $g$ to $e$ for a single frequency
$\omega_k$ is
\begin{eqnarray}
  \gamma_k = \frac{|c_e(t)|^2}{t} =
  \frac{\pi}{4}t\,\Omega_k^2\,D(\delta_k t/2),
\end{eqnarray}
where
\begin{eqnarray}
  D(x) \equiv \frac{\sin^2 x}{\pi x^2}.
\end{eqnarray}
The total decay rate is obtained by summing the decay rates for all
the fields in the comb:
\begin{eqnarray}
  \label {eqn:decay-rate-comb}
  \gamma = \sum_k \gamma_k =
  \frac{\pi}{4}t\sum_k \Omega_k^2\,D(\delta_k t/2).
\end{eqnarray}
To evaluate this expression we need to know the distribution of Rabi
frequencies $\Omega_k$.
This information can be obtained by forming a beat note between the
FORT and Raman beams on a photodetector, and measuring the power
spectrum $S(\omega)$ of the photocurrent using a spectrum analyzer.
Let us first consider this measurement for a monochromatic Raman beam,
and then generalize to a spectrally broad Raman beam.
If both the FORT and Raman beams are monochromatic, with optical
frequencies $\omega_F$ and $\omega_R$, then the resulting photocurrent
$i(t)$ is given by
\begin{eqnarray}
  i(t) =
  i_F + i_R + 2\eta \cos((\omega_F - \omega_R)t)\,\sqrt{i_F i_R},
\end{eqnarray}
where $i_F$ and $i_R$ are the cycle-averaged photocurrents for the
FORT and Raman beams taken individually and $\eta$ is the heterodyne
efficiency.
Thus, the power spectrum of the photocurrent has a spike at the
difference frequency $\Delta \equiv \omega_F - \omega_R$:
\begin{eqnarray}
  S_c(\omega) = P_c\,\delta(\omega - \Delta),
\end{eqnarray}
where the integrated power $P_c$ of the spike is proportional to
$i_F i_R$.
If the difference frequency $\Delta$ is tuned to Raman resonance
($\Delta = \omega_A$), then the FORT-Raman pair drives coherent Raman
transitions with a Rabi frequency $\Omega_c$ that is proportional to
$\sqrt{i_F i_R}$, so
\begin{eqnarray}
  \label{eqn:coherent-flops}
  \Omega_c^2 = \alpha P_c,
\end{eqnarray}
where $\alpha$ is a constant that depends on various calibration
factors.

Now consider the case of a spectrally broad Raman beam, which results
in a photocurrent with power spectrum $S_i(\omega)$.
The effective Rabi frequency $\Omega_k$ corresponding to comb line $k$
is given by
\begin{eqnarray}
  \Omega_k^2 = \alpha\,S_i(\omega_k)\,\delta\omega,
\end{eqnarray}
where $\delta\omega$ is the frequency spacing between adjacent comb
lines.
Substituting this result into equation (\ref{eqn:decay-rate-comb}),
and replacing the sum with an integral, we obtain
\begin{eqnarray}
  \gamma =
  \frac{\pi}{4}\alpha t \int S_i(\omega)\,
  D((\omega - \omega_A)t/2)\,d\omega.
\end{eqnarray}
If the power spectrum near $\omega_A$ is flat over a bandwidth $\sim
1/t$, then we can approximate $D$ as a delta function and perform the
integral:
\begin{eqnarray}
  \gamma =
  \frac{\pi}{2}\alpha\, S_i(\omega_A).
\end{eqnarray}
It is convenient to use equation (\ref{eqn:coherent-flops}) to
eliminate the calibration factor $\alpha$:
\begin{eqnarray}
  \gamma = \frac{\pi}{2}\frac{S_i(\omega_A)}{P_c}\,\Omega_c^2.
\end{eqnarray}
The spectrum analyzer trace given in Figure
\ref{fig:noise-spectrum}a displays the power spectrum in terms of
the power $P_i(\nu) \simeq 2\pi B\,S_i(\omega)$ in a bandwidth
$B = 3\,\mathrm{kHz}$, so we can also write this as
\begin{eqnarray}
  \gamma =
  \frac{1}{4}\frac{P_i(\omega_A/2\pi)}{P_c}\frac{\Omega_c^2}{B} =
  \frac{1}{4}(1-m^2/16)\,\frac{\Omega_0^2}{B}\,
  \frac{P_i((\Delta_{HF}+\omega_B\,m)/2\pi)}{P_c},
\end{eqnarray}
where we have substituted $\Omega_c = (1 - m^2/16)^{1/2}\,\Omega_0$
and $\omega_A = \Delta_{HF} + \omega_B\,m$.

We can calculate the time evolution of the atomic populations using
rate equations.
It is straightforward to show that the decay rate $e\rightarrow g$ is
also given by $\gamma$, and from the rate equations one can show that
the excited state population is
\begin{eqnarray}
  p_e(t) = \frac{1}{2}(1 - \exp(-2\gamma t)).
\end{eqnarray}

We can calculate the decay rates for the noise spectrum shown in
Figure \ref{fig:noise-spectrum}.
For this noise spectrum the power $P_i(\nu)$ has roughly the same value
$\bar{P}_i$ at the frequencies of all the $m \ne 0$ Zeeman
transitions, so we can write the decay rates for these transitions as
\begin{eqnarray}
  \gamma(|3,m\rangle \rightarrow |4,m\rangle) =
  \gamma(|4,m\rangle \rightarrow |3,m\rangle) =
  (1 - m^2/16)\,\Gamma,
\end{eqnarray}
where
\begin{eqnarray}
  \label{eqn:Gamma}
  \Gamma \equiv (1/4)(\Omega_0^2/B)(\bar{P}_i/P_c).
\end{eqnarray}
From the power spectrum for the noise shown in Figure
\ref{fig:noise-spectrum}a we have that
$\bar{P}_i = -63\,\mathrm{dBm}$, and
from the power spectrum for the coherent signal shown in Figure
\ref{fig:noise-spectrum}b we have that $P_c = -36\,\mathrm{dBm}$,
where the corresponding Rabi frequency is $\Omega_0 =
(2\pi)(120\,\mathrm{kHz})$.
Substituting these values into equation (\ref{eqn:Gamma}), we obtain
$\Gamma = 0.084\,\mathrm{\mu s}^{-1}$.

\end{document}